\documentclass[twocolumn, prl, showpacs]{revtex4}

\usepackage{epsfig}

\graphicspath{{Fig/}}

\usepackage{amsmath}
\usepackage{amsfonts}
\usepackage{amssymb}

\renewcommand{\natural}{{\vrule height .48em width .03em depth 0ex \kern -.30em {\rm N}}}

\begin{document}

\title{Finding instabilities in the community structure of complex networks}

\author{David Gfeller$^1$} 
\author{Jean-C\'edric Chappelier$^2$}
\author{Paolo De Los Rios$^1$}
\affiliation{$^1$ Laboratoire de Biophysique Statistique, SB/ITP and\\$^2$ School of Computer and Communication Sciences,\\Ecole Polytechnique F\'ed\'erale de Lausanne, CH-1015, Lausanne, Switzerland}

\date{\today}

\begin{abstract}
  The problem of finding clusters in complex networks has been extensively
  studied by mathematicians, computer scientists and, more recently,
  by physicists. Many of the existing algorithms partition a network
  into clear clusters, without overlap. We here introduce a method to
  identify the nodes lying ``between clusters'' and that allows for a
  general measure of the stability of the clusters. This is done by
  adding noise over the weights of the edges of the network.  Our method can in
  principle be applied with any clustering algorithm, provided that it
  works on weighted networks.  We present several applications on real-world
  networks using the Markov Clustering Algorithm (MCL).
\end{abstract}


\maketitle

The framework of complex networks provides a remarkable tool for the
analysis of complex systems consisting of many interacting entities
\cite{Albert2002, Newman2003-2}. Systems such as the Internet
\cite{Faloutsos1999}, the interaction map of proteins
\cite{Vazquez2003}, social networks \cite{Newman2003-1}, etc. have
been successfully described by considering them as complex networks.
Historically the first theoretical model to describe interacting complex systems was the Erd\"os-R\'enyi graph
\cite{Erdos1959}. However, this model fails to describe several
features observed frequently in real-world networks. The two most
famous ones are the degree distribution \cite{Barabasi1999} and the
clustering coefficient \cite{Watts1998}. Recently different models
have been proposed to give a more realistic understanding of those features.

\begin{figure}
\begin{center}
\rotatebox{0}{\resizebox{!}{4cm}{%
    \includegraphics[width=0.6\textwidth]{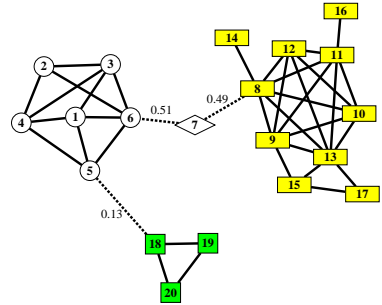}}}
\caption{Small toy network with one unstable node (7). The clusters obtained without noise are labeled with different colors. Only probabilities $p_{ij}<0.8$ are shown (dashed edges). $r=1.6$ and $\sigma=0.5$}
\label{net}
\end{center}
\end{figure}

Another characteristic of the topology of complex networks is their
cluster structure. In real-world networks, it is common to have small
sets of nodes highly connected with each other but with only a few
connections to the rest of the network. Finding the clusters of a
network is a crucial point in order to understand its internal
structure. A large amount of clustering algorithms have been
developed, each of them attempting to find a reasonably good partition
of the network \cite{Girvan2002, Capocci2004, Latapy2004, VanDongen2000, Radicchi2004}. In most of the cases those
algorithms partition the network into non-overlapping clusters,
assigning each node to a given cluster (``hard-clustering'').
However, the resulting clustering is sometimes questionable,
especially for nodes that ``lie on the border'' between two clusters.
We design such nodes as \emph{unstable} nodes. Fig.~\ref{net} shows a
typical case where a node (7) lies exactly between two clear clusters.

Defining and identifying unstable nodes is closely related to the
problem of evaluating the stability of the clustering. A first attempt
was proposed by Wilkinson \cite{Wilkinson2004} by modifying the
Girvan-Newman algorithm \cite{Girvan2002}. Recently several
non-deterministic clustering algorithms have been developed
\cite{Reichardt2004, Guimera2005, Duch2005}. Using the stochasticity
of the output, one can probe the stability of the clustering. In this work,
we introduce a general method to find unstable nodes and evaluate
the stability of the clusters. Instead of having a stochastic element in the algorithm, we propose to introduce stochasticity in the network itself and to use a hard-clustering algorithm (we chose the Markov Clustering Algorithm, MCL \cite{VanDongen2000}, but the method does not depend explicitly on this choice). The idea is to add a random noise over the weight of the edges of the network (in this study the noise added over the weight of the edges, initially equal to 1, is equally distributed between $[-\sigma, \sigma]$, $0<\sigma <1$). Noise in this context is not only a
useful tool to reveal cluster instabilities, but it has actually a
deeper interpretation. In many real-world networks, edges are often
provided with some intrinsic weights, but usually no information is
given about the uncertainties over these values. Adding some noise
could fill this lack, although arbitrarily, to take into account the
possible effects of uncertainties.

 Comparing how the clusters change from one noisy realization to another one provides informations that could not have been extracted with the standard clustering algorithms. For instance some nodes will ``switch from one cluster to another'' between different runs of the clustering algorithm with noise (nodes 7 in Fig.~\ref{net}).
Clusters are only determined by the nodes they are composed off. Hence,
what ``switch from one cluster to another'' means has to be
defined more precisely. We first introduce a probability $p_{ij}$ for
the edge between node $i$ and node $j$ of connecting two nodes in the
same cluster. After several runs of the clustering algorithm with the
noise, one obtains a network where edges with $p_{ij}=1$ are always
within a cluster and edges with a $p_{ij}$ close to 0 connect two
different clusters. Edges with a probability lower than a threshold
$\theta$ will be considered as external edges (typically $\theta=0.8$). By removing those
edges, one gets a disconnected network. Here, we use the
word \emph{cluster} for the clusters obtained without noise, and
\emph{subcomponent} for the disconnected parts of the network after
the removal of the external edges. If the community structure of the
network is stable under several repetitions of the clustering with
noise, the subcomponents of the disconnected network will correspond
to the clusters obtained without noise. In the opposite case a new
community structure will appear with some similarity with the initial
one. In order to identify which subcomponents correspond to the
initial clusters, we introduce the notion of similarity between two
sets of nodes. If $E_1$ (resp. $E_2$) is the set of clusters (resp.
the set of subcomponents), we use the following definition of the
similarity ($s_{ij}$) between cluster $C_{1j} \in E_1$ and subcomponent $C_{2i} \in E_2$:
\[s_{ij}=\frac{\vert C_{2i}\cap C_{1j}\vert }{\vert C_{2i}\cup C_{1j}\vert },\hspace{1mm} 1\leq i\leq \vert E_2\vert,\hspace{1mm} 1\leq j\leq \vert E_1\vert.\]
If $C_{1j}=C_{2i}$, $s_{ij}=1$ and if $C_{1j}\cap C_{2i}=\emptyset$, $s_{ij}=0$. For every $C_{1j} \in E_1$, we find the subcomponent $C_{2i}, 1\leq i\leq \vert E_2\vert$, with the maximal similarity and identify it with the cluster $C_{1j}$ (most of the time $C_{2i}$ corresponds to the stable core of the cluster $C_{1j}$). If there is more than one of such subcomponents, none of them will be identified with the cluster. In practice, this latter case is extremely rare.

For example, the network in Fig.~\ref{net} consists of three clusters (the three colors) and four subcomponents (\{1,2,3,4,5,6\}, \{7\}, \{8,9,10,11,12,13,14,15,16,17\}, \{18,19,20\}). Our method identifies the three biggest subcomponents with the three clusters, while the subcomponent \{7\} is not identified with any cluster. 

Nodes belonging to subcomponents that have never been identified with any cluster could be defined as unstable nodes. However in some cases a big cluster splits into two subcomponents of comparable size. Assuming that almost half of the nodes of the cluster are unstable is not realistic and one would rather define a new cluster. In practice, subcomponents of four nodes or more correspond often to a cluster not detected by the algorithm. We therefore define the unstable nodes as the nodes belonging to subcomponents that have not been identified with a cluster and whose size is smaller than 4.

Locally we can address the question of the stability of the clusters by looking at the probabilities of the edges inside each cluster and around a cluster. For instance if all edges inside the cluster have probability $p_{ij}=1$ and all edges connecting the cluster to its neighbors have probability $p_{ij}=0$, we can conclude that the cluster is very stable.

From a more global point of view, it is important to understand if the partition found by the clustering algorithm corresponds actually to a real cluster structure. We propose the entropy as a measure of the stability of the cluster structure. In first approximation, we assume that the $p_{ij}$ are independent of each other and we define the average {\em Clustering Entropy} (CE) per edge as:
\[S=\frac{-1}{m}\sum_{(i,j)}\{p_{ij}\log_2 p_{ij}+(1-p_{ij})\log_2(1-p_{ij})\},\]
where the sum is taken over all edges and $m$ is the total number of edges in the network. If the network is totally unstable (i.e. in the most extreme case $p_{ij}=\frac{1}{2}$ for all edges), $S=1$, while if the edges are perfectly stable under noise ($p_{ij}=0$ or $1$), $S=0$.

The value of $S$ depends on the noise $\sigma$. Nevertheless it allows
for comparing with a network without predefined cluster structure. To
avoid biasing the comparison, we shall always compare the CE of a network with the one of a randomized version of the network
in which the degree of each node is conserved \cite{Maslov2002,
  Eriksen2003}, using the same $\sigma$. The randomized network plays the role of a null-model since the initial clusters (if present) are destroyed by the
rewiring process. Note however that we do not assume the randomized
network to have no apparent community structure \cite{Guimera2004}. If the
difference between the CE of the original network and the randomized
one is important (i.e. is not within the standard deviation from
different randomized versions), it shows that the network has an
internal cluster structure that differs fundamentally in terms of
stability from a network where the nodes have been connected randomly.

Before showing applications of our method to study the stability of the clusters, we briefly describe MCL \cite{VanDongen2000} that we used as a clustering technique. MCL is based on the
idea that when \emph{a random walk on a network visits a dense
  cluster, it will likely not leave it until many of its vertices have
  been visited}. However the idea of performing a random walk on a
network does not immediately lead to the clusters, since as the time
increases, the random walk will end up leaving one cluster for another. MCL
favors the most probable random walks, already after a small number of
steps, thereby increasing the probability of staying in the initial
cluster. The algorithm works as follows: 1)~take the adjacency matrix
$A$ of the network; add the self-edges (1's on the diagonal) and
normalize each column of the matrix to one, in order to obtain a
stochastic matrix $W$; 2)~take the $k^{th}$ power of the matrix $W$,
$k\in \natural$ (we used $k=2$); 3)~take the
$r^{th}$ power of every element of $W^k$ (typically $r\approx 1.5-2)$
and normalize each column to one; 4)~go back to~2).
After several iterations MCL converges to a matrix idempotent under
step 2) and 3). Only a few lines of the matrix have some non zero
entries that give the cluster structure of the network.  Note that the
parameter $r$ can tune the granularity of the clustering. A small $r$
corresponds to a few big clusters, whereas a big $r$ to smaller ones.

\begin{figure}
  \centering
  \rotatebox{270}{\resizebox{!}{6cm}{%
      \includegraphics[width=0.8\textwidth]{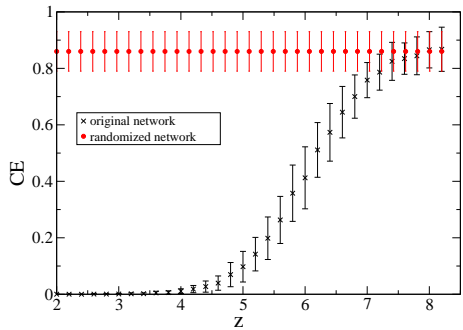}}}
  \caption{CE as a function of $z$, the average number of edges connecting a node from a given cluster to nodes of other clusters, for a network with 4 communities of 32 nodes. The error bars represent the standard deviation for different networks with the same $p_{in}$ and $p_{out}$. $r=1.85$, $\sigma=0.5$}
  \label{bloc}
\end{figure}

To illustrate the principle of the comparison based on the CE, we apply it on the well-known benchmark network introduced first in \cite{Girvan2002}. The network consists of 4 communities of 32 nodes. The nodes are connected with a probability $p_{in}$ if they belong to the same community and $p_{out}$ if not. Typically one chooses to vary $p_{in}$ and $p_{out}$ keeping the average degree of the nodes constant. In Fig.~\ref{bloc} we plot the CE of the network. $z$ is the average number of edges connecting a node from a given cluster to nodes of other clusters ($z=96\cdot p_{out}$). The average total degree is fixed at 16. When $z$ is small the clusters are very well defined and most of the algorithms correctly identify them. As $z$ increases, the clusters become more and more fuzzy and for $z>7$ even the best currently available algorithms fail to recover the exact cluster structure of the network (actually the cluster structure tends to disappear from the network). This corresponds to the point from which the comparison of the CE does not allow to differentiate between the network and a randomized network. We stress that the clustering entropy does not make reference to the assumed partition of the network into four clusters that, given the statistical nature of the links, cannot be guaranteed for every realization. It is thus an objective measure of the stability of the network under clustering.

Let us now turn to real-world networks. As a first example, we consider the ``karate club network'' built by
Zachary \cite{Zachary1977}. MCL correctly identifies the two
communities, which correspond to the actual division of the club. The
only unstable node is represented with a diamond. This node is
connected to four nodes of one community and five of the other one.
From a topological point of view, it is absolutely justified to
consider it as an unstable node. The CE of the network is 0.14. The
randomized network has an average CE of $0.27\pm0.1$ (average and
standard deviation of 100 randomized versions). Thus on average the CE is
significantly larger for the randomized network.

\begin{figure}
  \centering
  \rotatebox{0}{\resizebox{!}{4cm}{%
      \includegraphics[width=0.6\textwidth]{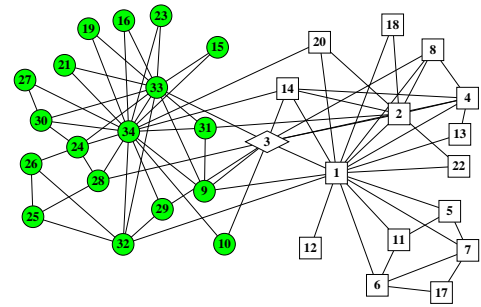}}}
  \caption{Zachary's karate club network. The two clusters are represented with two different colors. The unstable node is represented by a diamond. $r=1.8$, $\sigma=0.5$}
  \label{karate}
\end{figure}

\begin{figure}[b]
  \centering
  \rotatebox{270}{\resizebox{!}{6cm}{%
      \includegraphics[width=0.6\textwidth]{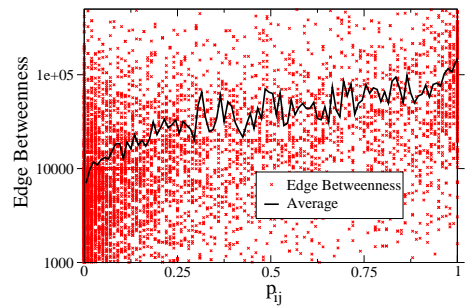}}}
  \caption{Edge betweenness versus $p_{ij}$ for a component of 9997
    nodes from the synonymy network. $r=1.6$, $\sigma=0.5$}
  \label{tebet}
\end{figure}

\begin{figure}
  \centering
  \rotatebox{270}{\resizebox{!}{6cm}{
  \includegraphics[width=0.6\textwidth]{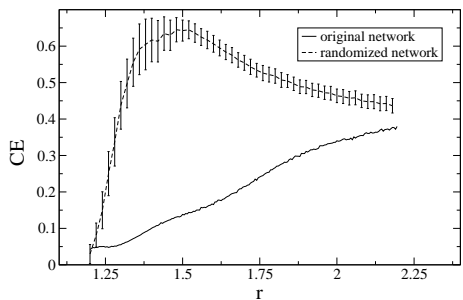}}}
  \caption{CE as a function of the parameter $r$ for a network of 185 nodes. The dashed curve is the average over 50 randomized versions and the error bars correspond to the standard deviation. $\sigma=0.5$.}
  \label{lentropy}
\end{figure}

We studied a linguistic network based on the relation of synonymy in
French \cite{Gfelleretal2005a}. The nodes are the words in a given
sense. Two nodes are connected if they are considered as synonyms. We
applied MCL on the larger disconnected components of the network (up
to 10000 nodes) and found a much better lexical representation of the
synonyms. The natural interpretation of unstable nodes in the case of
a synonymy network is that they correspond to ambiguous words. As a
validation of our results, we can measure the clustering coefficient
of the unstable nodes. Averaging over the whole network, we have a
clustering coefficient of 0.26 for the unstable nodes and 0.45 for the
stable nodes. Furthermore the betweenness \cite{Newman2001} of
unstable nodes is on average 1.6 times larger.  The important
difference was expected since unstable nodes often lie between
clusters, and therefore usually do not have a large clustering
coefficient, but have larger betweenness. Moreover the plot of the
edge betweenness versus the probability $p_{ij}$ shows that external
edges have on average a larger betweenness (Fig.~\ref{tebet}), which is consitent with the Girvan-Newman clustering algorithm \cite{Girvan2002}.
Fig.~\ref{lentropy} shows how the CE varies with the parameter $r$ of MCL for a component of 185 nodes compared with a randomized version of the same component. For $1.3<r<2$, the difference of behavior is striking. This shows that the clusters are not a by-product of the clustering algorithm, but correspond to a real community structure of the network. 

We finally applied MCL on the protein folding network of the anti-parallel $\beta$-sheet peptide developed by Rao and Caflisch~\cite{Rao2004}. The network consists of almost 80000 nodes. MCL correctly identifies the native state (or at least part of it) and other stable configurations such as the curl-like trap. Studying the stability of the clusters, we restricted ourselves to the network with cut-off (1069 nodes). We can again compare the CE. For $r=1.6$ and $\sigma=0.5$ we have a CE of $0.12$, while the randomized network shows an entropy of $0.3\pm 0.05$ (average over 50 randomized versions).

Note that the parameter $\sigma$ can in principle influence the results. With $\sigma$ close to 0, we cannot detect the unstable nodes, while with $\sigma$ bigger than one, the topology of the network changes dramatically. However the results do not change significantly for a broad range of values of $\sigma$ around $0.5$.  For instance in the network displayed in Fig.~\ref{net}, the node 7 was identified as the only unstable node for $0.15\leq \sigma \leq 0.8$. Moreover very similar results are obtained using a gaussian distribution for the noise.

In conclusion, the introduction of the noise on the edges and the probabilities $p_{ij}$ provides a well-defined and objective way to identify unstable nodes and to deal with ambiguities in clustering. The method performs well on the small test networks presented above. As a validation of our results for larger networks that can hardly be visualized, we have seen that the clustering coefficient of the unstable nodes is usually much lower than the average clustering coefficient of the whole network. Moreover these nodes have, on average, a larger betweenness, which is also expected for nodes lying between clusters. Nevertheless we could not have identified the unstable nodes only by comparing the clustering coefficient and the betweenness since very stable nodes may still have a large betweenness and a small clustering coefficient, and vice versa.
The Clustering Entropy allows for a quantitative comparison between a network and a null-model. We have found that in many examples the difference was clear, assuring that the clusters detected by MCL are neither the result of random fluctuations in the modularity of the network \cite{Guimera2004}, nor an artefact of the clustering algorithm. Finally, since the method does not depend on a particular clustering algorithm, it can in principle be implemented using any other clustering technique than MCL we used here.

This work was financially supported by the EU
commission by contracts COSIN (FET Open IST 2001-33555) and DELIS (FET
Open 001907) and the OFES-Bern under Contract No. 02.0234.
We are thankful to M.E.J. Newman for providing the data of the
Zachary's karate club network and to F. Rao and A. Caflisch for the
protein folding network. 

\bibliography{bibcondmat2005}

\end{document}